# A promotion for odd symmetric discrete Fourier transform

Rui Li

*Abstract*—DFT is the numerical implementation of Fourier transform (FT), and it has many forms. Ordinary DFT (ODFT) and symmetric DFT (SDFT) are the two main forms of DFT. The most widely used DFT is ODFT, and the phase spectrum of this form is widely used in engineering applications. However, it is found ODFT has the problem of phase aliasing. Moreover, ODFT does not have many FT properties, such as symmetry, integration, and interpolation. When compared with ODFT, SDFT has more FT properties. Theoretically, the more properties a transformation has, the wider its application range. Hence, SDFT is more suitable as the discrete form of FT. In order to promote SDFT, the unique nature of SDFT is demonstrated. The time-domain of even-point SDFT is not symmetric to zero, and the author corrects it in this study. The author raises a new issue that should the signal length be odd or even when performing SDFT. The answer is odd. However, scientists and engineers are accustomed to using even-numbered sequences. At the end of this study, the reasons why the author advocates odd SDFT are given. Besides, even sampling function, discrete frequency Fourier transform, and the Gibbs phenomenon of the SDFT are introduced.

*Index Terms*—Discrete frequency Fourier transform, even sampling function, integral property, phase spectrum, symmetry properties, symmetric DFT

## I. Introduction

Discrete Fourier transform (DFT) was raised in diverse settings and is used by practitioners in diverse fields that it appears in many different forms [1]. Anyone who regularly works with DFT will eventually encounter it in more than one form. Discreteness and ingenuity must be utilized to ensure that the input is in the proper form and the output is interpreted correctly [1]. More information can be referred to Reference [2].

DFT has two main forms: the ordinary form and the symmetric form. The most widely used DFT is ordinary DFT (ODFT), and the well-known FFT [3] is the fast algorithm of ODFT. With the help of FFT, the computation time is much shorter, and the memory cost is much lesser. Under the promotion of FFT, the ordinary form is renowned in digital signal processing, and its impact on contemporary society is enormous and unprecedented. Researchers have paid much attention to this form, and this form is widely used in mechanical structure fault diagnosis [4], sonar and radar detection, spectroscopy[5], seismic location positioning, celestial mechanics research[6], and so on. One important DFT is symmetric DFT (SDFT) [1], also known as unaliased DFT [7] or centered DFT [8]. This form is normally used in interpolation, data compression, and noise removal [9].

SDFT and ODFT are orthogonal transforms. In one dimensional transform, the amplitude spectrum of the two transforms is the same, whereas the phase spectrum is different. Phase spectrum is widely used in engineering applications, such as optical flow [10], video motion magnification [11], [12], frequency estimation [13]–[15], and video frame interpolation [16]. Nowadays, phase spectrum becomes increasingly important in image processing [17]–[19]. According to the knowledge of the author, those applications are based on ODFT. However, it is found ODFT has the problem of phase aliasing. If one applies SDFT to those applications, the output results would be different. Although there is no sign that there will be better results, there is also no sign that there will be bad results. Hence, the study of SDFT is of significance, and its huge potentiality makes SDFT are too attractive to neglect.

DFT is the numerical implementation of Fourier transform (FT). Based on the following four reasons, the author recommends SDFT when performing FT. Firstly, the time-domains of FT and SDFT are symmetric to zero, whereas the time-domain of ODFT is not symmetric to zero. Secondly, if one turns the signal head around, ODFT gets two completely different spectra, whereas FT and SDFT get a pair of the conjugate spectrum. Thirdly, ODFT does not have the same symmetric properties as FT, whereas SDFT has these properties. Lastly, according to Noether's theorem [20], symmetry seems to be the prerequisite of a differentiable physical system with the conservation law.

According to the parity of signal length, SDFT is divided into odd SDFT and even SDFT. However, the time interval of even SDFT is not strictly symmetrical to zero [1]. Moreover, the author finds that this unsymmetrical problem has not been solved yet [9]. In this study, the author corrects even SDFT. After correction, the window function, sampling function, complex orthogonal basis, transform matrix, and phase spectrum is different. The differences are introduced in this study. Besides, some unique properties of SDFT are exhibited, as well as its derivation. For example, the symmetry properties and the integral properties.

Zero-padding is a technique of increasing spectrum samples. The density of spectrum samples increased after padding a large

School of Mechanical Science and Engineering, Huazhong University of Science and Technology, Wuhan, 430074, China (e-mail: 824089827@qq.com or jindui123@126.com).

number of zeros to the input signal. The zero-padding technique is not limited to the time-domain; it is also applicable in the frequency-domain. That is to say, frequency-domain zero-padding can be used to increase the number of time-domain samples. Corresponding to discrete-time frequency transform (DTFT), this study proposes the discrete frequency Fourier transform (DFFT).

An interesting issue is that: should the signal length be odd or even when performing DFT. Scientists and engineers are accustomed to using even-numbered signals. However, the advisable length is odd according to the author's analysis. Reasons why the author advocates odd SDFT are given at the end of this study.

The rest of the manuscript is structured as follows. In section II, the background of this study is introduced. The correction for even SDFT is presented in section III. The unique properties of SDFT are exhibited in section IV. The zero-padding technique is discussed in section V. Reasons that the author recommends odd SDFT are exhibited in section VI.

## II. BACKGROUND

*A. The asymmetry problem of SDFT*

The mathematical formula of SDFT [1], [7], [9] is related to the parity of signal length $N$. When $N$ is odd ($N=2k+1$), and the mathematical formula is (1). When $N$ is even ($N=2k$), the mathematical formula is (2). One characteristic of SDFT is $n$ (represents time) are integers. That is to say, odd SDFT and even SDFT have the same sampling function. It is easy to find the time interval of even SDFT is not symmetric to zero. The asymmetry problem of even SDFT has not been solved yet [9].

$$X(m) = \sum_{n=-k}^{k} x(n)e^{-i2\pi mn/N} \qquad (1)$$

$$X(m) = \sum_{n=-k}^{k-1} x(n)e^{-i2\pi mn/N} \qquad (2)$$

*B. Phase aliasing of shifted window*

DFT of a discrete sequence is the FT's convolution of the signal, window, and sampling function. Hence, window function plays an important role in DFT spectrum analysis. The symmetrical rectangular window is shown in Fig. 1 (a). Assuming the sampling frequency is $f_s$, and $N$ samples are obtained, the time value range is $t\in(-N/f_s/2, N/f_s/2)$. The FT of the symmetrical window is:

$$W(f) = \int_{-\infty}^{\infty} w(t)e^{-i2\pi ft}dt = \frac{N}{f_s}\text{sinc}(Nf/f_s). \qquad (3)$$

The window of ODFT refers to Fig. 1 (b). Compared with the symmetrical window, this window is shifted, and the distance it shifts is $(N-1)/2$. According to the translation property of FT, the FT of the shifted window is

$$W_o(f) = W(f)e^{-\frac{i2\pi f(N-1)/2}{N}} = \frac{N}{f_s}\text{sinc}(Nf/f_s)e^{ic_1 f}. \qquad (4)$$

When comparing the FT of the symmetrical window and the shifted window, the phase spectrum of the shifted window is a linear function of frequency. This phenomenon is named phase aliasing.

The value of constant $c_1$ is $-\pi(N-1)/N$, which is the same as Eq. (20) in reference [21] and Eq. (3) in reference [22]. In other circumstances, the value is $-\pi$, referring to Eq. (8) in reference [23], Eq. (7) in reference [24], and Eqs. (22) and (23) in reference [25]. The cause of this phenomenon is that the initial value of the window function. If the initial value is not zero, the constant is $-\pi(N-1)/N$. If the initial value is zero, then the constant is $-\pi$. For example, the constant of the rectangular and symmetric Hann window is $-\pi(N-1)/N$, and the constant of the periodic Hann window is $-\pi$.

The zero-point of SDFT locates in the middle of a signal theoretically. However, the zero-point of even SDFT is not in the middle, as shown in Fig. 1 (c). The window shift of even SDFT is 1/2, then Eq. (4) becomes

$$X_{ws}(f) = X_w(f)e^{ic_2 f}. \qquad (5)$$

The value of constant $c_2$ is $\pi/N$. The asymmetric window complicates the SDFT spectrum, and it brings a lot of trouble to discrete spectrum analysis.

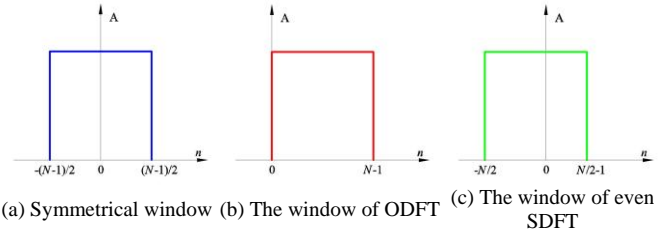

(a) Symmetrical window (b) The window of ODFT (c) The window of even SDFT

Fig. 1. Schematic diagram of three windows

*C. Complex orthogonal basis*

A set of complex orthogonal bases in Euclidean space $\mathbf{C}^{N\times N}$ has $N$ orthogonal vectors, and these $N$ vectors compose an orthogonal transform matrix. If one changes the $N$ orthogonal vectors' arrangement order, it still composes an orthogonal transform matrix. Theoretically, there are many orthogonal transforms, and they have the same orthogonal basis. ODFT and SDFT share a complex orthogonal basis. The difference between the two transform matrices is the arrangement order of the $N$ complex orthogonal vectors. The asymmetric problem cannot be solved by re-ordering the complex orthogonal vectors.

The ODFT of signal $x(n)$ is defined as:

$$X_o(m) = \sum_{n=0}^{N-1} x(n)e^{-i2\pi mn/N}. \qquad (6)$$

Where $\pi$ is the circumference rate, e is the Euler's number, i is the imaginary unit, $m$ represents frequency, and $n$ represents time. The value range of $m$ is $\{m\in\mathbf{N}|0\le m\le N-1\}$.

The ODFT and IDFT matrix are $N$-by-$N$ complex matrix, and they are the focus of this section. The transform matrix of ODFT is $\boldsymbol{D}_{\text{forward}} = [\boldsymbol{q}(0), \boldsymbol{q}(1), \boldsymbol{q}(2), \cdots, \boldsymbol{q}(n), \cdots, \boldsymbol{q}(N-1)]$, and the inverse matrix is $\boldsymbol{D}_{\text{inverse}}= [\boldsymbol{q}(0), \boldsymbol{q}(-1), \boldsymbol{q}(-2), \cdots, \boldsymbol{q}(-n), \cdots, \boldsymbol{q}(-N+1)]$, in which $\boldsymbol{q}(n)$ ($\boldsymbol{q}\in\mathbf{C}^{N\times N}$) are perpendicular complex vectors, and it can be written as:

$$\boldsymbol{q}(n) = \left[e^{-0i}, e^{-\frac{i2\pi n}{N}}, e^{-\frac{i2\pi 2n}{N}}, \cdots, e^{-\frac{i2\pi(N-1)n}{N}}\right]^T. \qquad (7)$$

Where T represents the transposition, and the value range of $n$ is $\{n\in\mathbf{Z}|\ 0\le n\le N-1\}$. The Hadamard product of arbitrary vector $\boldsymbol{q}(n)$ with all one vector is itself, and this property can be used

for manipulating those complex vectors. Assuming all one vector is $o(n)$, then we have

$$q(n) = q(n) * o(n). \tag{8}$$

For arbitrary integer $m$, $e^{(i2\pi m)}$ is equal to 1, then $o(n)$ can be written as

$$o(n) = [e^{0i}, e^{i2\pi}, e^{i2\pi 2}, \cdots, e^{i2\pi m}, \cdots, e^{i2\pi(N-1)}]^T. \tag{9}$$

Substituting (7) and (9) into (8), then we have

$$q(n) = q(n - N). \tag{10}$$

It is easy to deduce that $q(1) = q(-N+1)$, $q(2) = q(-N+2)$, $\cdots$, $q(n) = q(-N+n)$, $\cdots$, $q(N-1) = q(-1)$. We may conclude that the orthogonal basis of ODFT and IDFT is the same. The difference between them is the arrangement order of those $N$ perpendicular complex vectors.

The complex orthogonal basis of even SDFT is $[q(-N/2), q(-N/2+1), \cdots, q(0), \cdots, q(N/2-2), q(N/2-1)]$. According to (10), the complex orthogonal basis of even SDFT is the same as that of ODFT. The complex orthogonal basis is linked with the sampling function strongly. Both ODFT and SDFT use odd sampling functions, which are the foundation for the same complex orthogonal basis.

TABLE I
VARIOUS EQUIDISTANT SAMPLING FUNCTION

| Function | Time domain ($s(t)$) | Frequency domain ($S(f)$) |
|---|---|---|
| Odd | $\sum_{n=-\infty}^{\infty} \delta(t - n\Delta T)$ | $f_s \sum_{j=-\infty}^{\infty} \delta(f - jf_s)$ |
| Even | $\sum_{n=-\infty}^{\infty} \delta(t - (n+0.5)\Delta T)$ | $f_s \sum_{j=-\infty}^{\infty} (-1)^j \delta(f - jf_s)$ |
| Generalized | $\sum_{n=-\infty}^{\infty} \delta(t - (n+r)\Delta T)$ | $f_s e^{-i2\pi f r/f_s} \sum_{j=-\infty}^{\infty} \delta(f - jf_s)$ |
| Reversal | $\begin{cases} \delta(t) & t = 2n\Delta T \\ -\delta(t) & t = (2n+1)\Delta T \\ 0 & \text{else} \end{cases}$ | $f_s \sum_{j=-\infty}^{\infty} \delta(f - (j+0.5)f_s)$ |

Where $\Delta T$ is the time interval between two samples, $f_s$ is the sampling frequency, $\delta$ is the dirichlet function. The value range of time is $\{t \in \mathbf{R} | -\infty < t < \infty\}$, the value range of frequency is $\{f \in \mathbf{R} | -\infty < f < \infty\}$, the value range of $n$ is $\{n \in \mathbf{N} | -\infty < n < \infty\}$, the value range of $r$ is $\{r \in \mathbf{R} | -0.5 < r \leq 0.5\}$, the value range of $j$ is $\{j \in \mathbf{N} | -\infty < j < \infty\}$.

## III. THE CORRECTION

### A. FT of the even sampling function

Assuming the sampling frequency is $f_s$, then the time interval between arbitrary two samples is $\Delta T = 1/f_s$. Fig. 2 (a) plots the most frequently used sampling function (the odd sampling function), and the definition of it is

$$s_o(t) = \sum_{n=-\infty}^{\infty} \delta(t - n\Delta T). \tag{11}$$

In which $n$ is an integer, and $\delta$ is the Dirichlet function. The FT of the odd sampling function is plotted in Fig. 2 (b). Fig. 2 (c) plots the even sampling function, which is an optional sampling function. The definition of it is

$$s_e(t) = \sum_{n=-\infty}^{\infty} \delta(t - (n+0.5)\Delta T). \tag{12}$$

The two sampling functions are relevant, and the relationship between the two sampling functions is

$$s_e(t) = s_o(t - \Delta T/2) \text{ or } s_e(t) = s_o(t + \Delta T/2). \tag{13}$$

The FT of the odd sampling function is

$$S_o(f) = f_s \sum_{j=-\infty}^{\infty} \delta(f - jf_s). \tag{14}$$

According to the translation property, the FT of the even sampling function is derivable from the odd sampling function. For forward translation, according to the translation property, the FT of even sampling function is

$$S_e(f) = f_s e^{i2\pi f/(2f_s)} \sum_{j=-\infty}^{\infty} \delta(f - jf_s). \tag{15}$$

For backward translation, the FT of even sampling function is

$$S_e(f) = f_s e^{-i2\pi f/(2f_s)} \sum_{j=-\infty}^{\infty} \delta(f - jf_s). \tag{16}$$

Equation (15) is equal to (16), the FT of even sampling function is (17), as shown in Fig. 2 (d).

$$S_e(f) = f_s \sum_{j=-\infty}^{\infty} (-1)^j \delta(f - jf_s) \tag{17}$$

### B. Derivation validation of the even sampling function

In this sub-section, the author introduces another derivation of the FT of the even sampling function. The basic theory is that an odd sampling function can be decomposed into an odd and even sampling function with the same frequency. According to the linearity of FT, the FT of even sampling function is derivable from the odd sampling function.

$$s_{2o}(t) = \sum_{n=-\infty}^{\infty} \delta(t - n\Delta T/2) \tag{18}$$

If we double the frequency of the odd sampling function, as shown in (18) or Fig. 2 (e), and according to (14), the FT of the frequency-doubled odd sampling function is (19), as shown in Fig. 2 (f).

$$S_{2o}(f) = 2f_s \sum_{j=-\infty}^{\infty} \delta(f - j2f_s) \tag{19}$$

Subtracting (11) from (18), we get the even sampling function, as shown in Fig. 2 (g) to Fig. 2 (h). According to the linearity of FT, the FT of the even sampling function is

$$S_e(f)' = S_{2o}(f) - S_o(f). \tag{20}$$

Substituting (19) and (14) into (20), that is to say, Fig. 2 (f) minus Fig. 2 (b), then we have

$$S_e(f)' = f_s \sum_{j=-\infty}^{\infty} (-1)^j \delta(f - jf_s). \tag{21}$$

Comparing (21) with (17), we can see they are the same. The sampling functions discussed above are equidistant, and a summary of equidistant sampling functions is listed in Table 1.

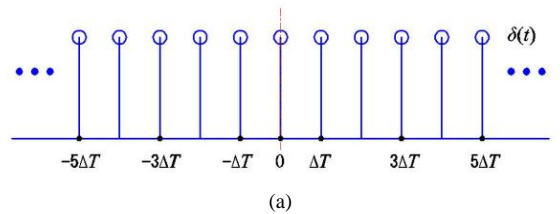

(a)

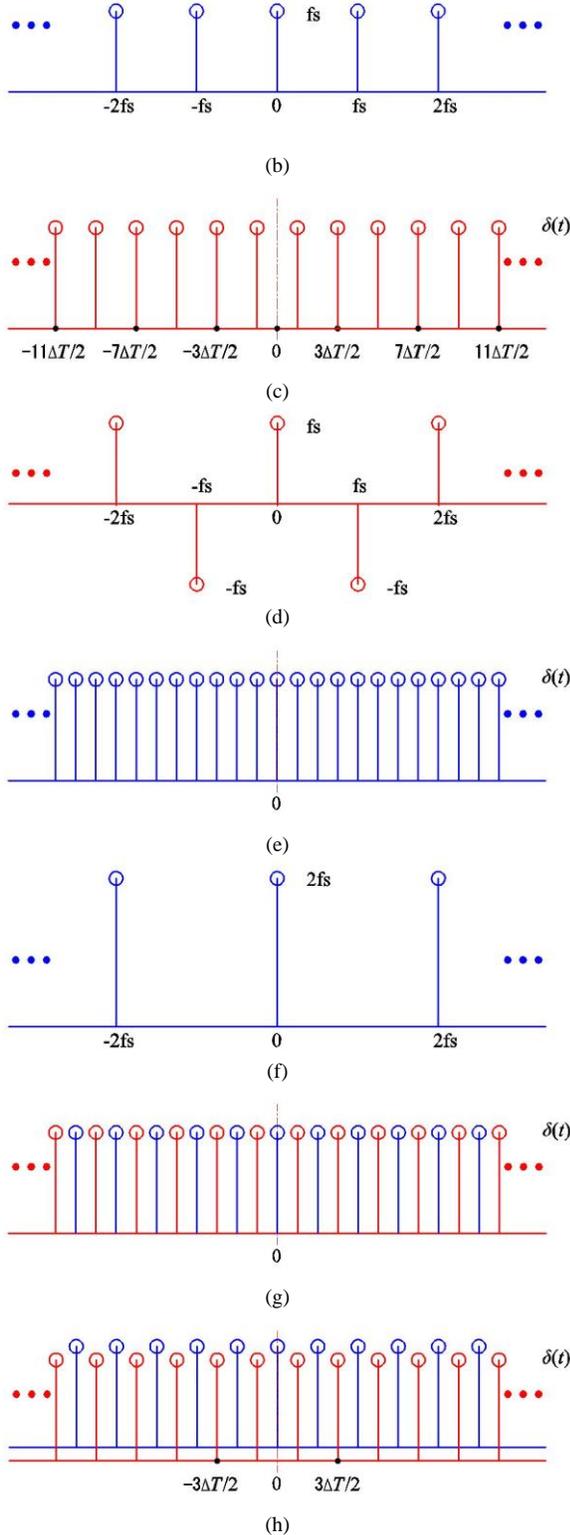

Fig. 2. Relationship between odd sampling and even sampling function. An odd sampling function can be decomposed into an odd sampling function and an even sampling function with the same frequency. Subplot (a) is an odd sampling function; Subplot (b) is the FT of the odd sampling function; Subplot (c) is an even sampling function; Subplot (d) is the FT of the even sampling function; Subplot (e) is a frequency-doubled odd sampling function; Subplot (f) is the FT of the frequency-doubled odd sampling function; Subplot (g) is the diagram of odd sampling function decomposition; Subplot (h) is the progress of the decomposition.

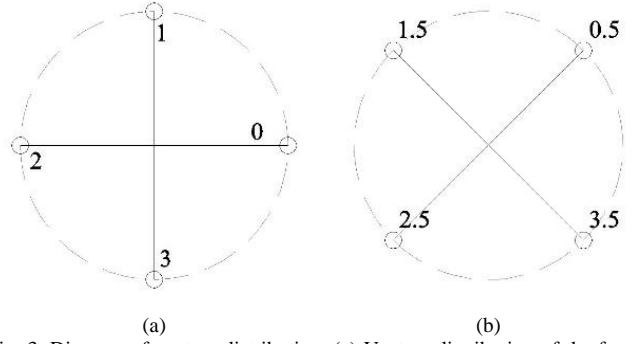

(a)           (b)

Fig. 3. Diagram of vectors distribution. (a) Vectors distribution of the former complex orthogonal basis; (b) Vectors distribution of the new basis.

### C. New complex orthogonal basis

The incorrect selection of sampling function leads to the asymmetry problem of even SDFT. In this section, the author corrects the even SDFT with the even sampling function. Hence, the basis of the corrected SDFT is different from the former basis.

Assuming $N$ is an even number, the new basis is $[q(0.5), q(1.5), q(2.5), \ldots, q(n+0.5), \ldots, q(N-0.5)]$, whereas the former basis is $[q(0), q(1), q(2), \ldots, q(n), \ldots, q(N-1)]$. The distribution of those perpendicular vectors refers to Fig. 3. Subplot (a) is the former basis, and subplot (b) is the new basis.

According to (7), the perpendicular complex vector $q(n+0.5)$ in the new basis is

$$q(n+0.5) = [e^{-0i}, e^{-\frac{i2\pi(n+0.5)}{N}}, e^{-\frac{i2\pi 2(n+0.5)}{N}}, \cdots, \\ e^{-\frac{i2\pi m(n+0.5)}{N}}, \cdots, e^{-\frac{i2\pi(N-1)(n+0.5)}{N}}]^T. \quad (22)$$

Where the value range of $n$ is $\{n \in \mathbf{N} | 0 \leq n \leq N-1\}$, according to the law of a standard complex orthogonal basis, the dot product of two arbitrary complex vectors in the new basis must satisfy the following constraints.

$$q(i+0.5) \cdot q(j+0.5) = \begin{cases} 0, & i \neq j \\ 1, & i = j \end{cases} \quad (23)$$

Substituting (22) into (23), the dot product of two arbitrary complex vectors is

$$q(i+0.5) \cdot q(j+0.5) = q(j+0.5)^H q(i+0.5) \\ = q(j)^H q(i) = q(i) \cdot q(j). \quad (24)$$

In which "H" represents the Hermitian operator. As introduced above, $q(i)$ and $q(j)$ are the perpendicular orthogonal vectors of the former complex orthogonal basis, and they all satisfy the following constraints.

$$q(i) \cdot q(j) = \begin{cases} 0, & i \neq j \\ N, & i = j \end{cases} \quad (25)$$

We may conclude that $[q(0.5), q(1.5), q(2.5), \ldots, q(n+0.5), \ldots, q(N-1+0.5)]$ compose a complex orthogonal basis, and the length of each perpendicular vector is $\sqrt{N}$. Divide each complex perpendicular vector by $\sqrt{N}$, and we obtain the standard complex orthogonal basis.

According to the derivation above, there are many complex orthogonal bases in $\mathbf{C}^{N \times N}$. For an arbitrary real number $r$, $[q(r), q(r+1), q(r+2), \ldots, q(r+n), \ldots, q(r+N-1)]$ compose a set of complex orthogonal basis. The derivation of this complex orthogonal basis is the same as that of $[q(0.5), q(1.5), q(2.5), \ldots, q(n+0.5), \ldots, q(N-1+0.5)]$. When $r$ is an integer, this complex orthogonal basis is the same as the former. When $r$ is not an integer, this complex orthogonal basis differs from the former.

The value of *m* that in (22) does not necessarily be an integer. That is to say, arbitrary *N* consecutive real numbers with unit interval satisfy (25), then (22) becomes (26). The proof is similar to the above proof that the author neglect.

$$\boldsymbol{q}(r) = [e^{-\frac{i2\pi hr}{N}}, e^{-\frac{i2\pi(h+1)r}{N}}, e^{-\frac{i2\pi(h+2)r}{N}}, \cdots, \\ e^{-\frac{i2\pi(h+m)r}{N}}, \cdots, e^{-\frac{i2\pi(h+N-1)r}{N}}]^T. \quad (26)$$

*D. Example of new complex orthogonal basis*

In four points ODFT, the four perpendicular orthogonal vectors are $\boldsymbol{q}(0)$, $\boldsymbol{q}(1)$, $\boldsymbol{q}(2)$, and $\boldsymbol{q}(3)$, as shown in (27). The orthogonality of those four vectors has been proven in various monographs and open courses.

$$\begin{aligned}\boldsymbol{q}(0) &= [1, 1, 1, 1]^T \\ \boldsymbol{q}(1) &= [1, -i, -1, i]^T \\ \boldsymbol{q}(2) &= [1, -1, 1, -1]^T \\ \boldsymbol{q}(3) &= [1, i, -1, -i]^T\end{aligned} \quad (27)$$

The constructed complex orthogonal bases depend on the selected frequency. When the selected frequencies are (-*N*/2: *N*/2-1), the four new perpendicular orthogonal vectors are listed in (28).

$$\begin{aligned}\boldsymbol{q}(0.5) &= \left[e^{-0i}, e^{-\frac{i\pi}{4}}, e^{-\frac{i\pi}{2}}, e^{-\frac{i3\pi}{4}}\right]^T \\ \boldsymbol{q}(1.5) &= \left[e^{-0i}, e^{-\frac{i3\pi}{4}}, e^{\frac{i\pi}{2}}, e^{-\frac{i\pi}{4}}\right]^T \\ \boldsymbol{q}(2.5) &= \left[e^{-0i}, e^{\frac{i3\pi}{4}}, e^{-\frac{i\pi}{2}}, e^{\frac{i\pi}{4}}\right]^T \\ \boldsymbol{q}(3.5) &= \left[e^{-0i}, e^{\frac{i\pi}{4}}, e^{\frac{i\pi}{2}}, e^{\frac{i3\pi}{4}}\right]^T\end{aligned} \quad (28)$$

The inner product of arbitrary two perpendicular vectors is zero, and the length of each perpendicular vector is 2, as shown in (29).

$$\begin{aligned}\boldsymbol{q}(0.5) \cdot \boldsymbol{q}(0.5) &= 1 \times 1 + e^{\frac{i\pi}{4}} \times e^{-\frac{i\pi}{4}} + e^{\frac{i\pi}{2}} \times e^{-\frac{i\pi}{2}} \\ &\quad + e^{\frac{i3\pi}{4}} \times e^{-\frac{i3\pi}{4}} = 4 \\ \boldsymbol{q}(1.5) \cdot \boldsymbol{q}(1.5) &= 1 \times 1 + e^{\frac{i3\pi}{4}} \times e^{-\frac{i3\pi}{4}} + e^{-\frac{i\pi}{2}} \times e^{\frac{i\pi}{2}} \\ &\quad + e^{\frac{i\pi}{4}} \times e^{-\frac{i\pi}{4}} = 4 \\ \boldsymbol{q}(2.5) \cdot \boldsymbol{q}(2.5) &= 1 \times 1 + e^{-\frac{i3\pi}{4}} \times e^{\frac{i3\pi}{4}} + e^{\frac{i\pi}{2}} \times e^{-\frac{i\pi}{2}} \\ &\quad + e^{-\frac{i\pi}{4}} \times e^{\frac{i\pi}{4}} = 4 \\ \boldsymbol{q}(3.5) \cdot \boldsymbol{q}(3.5) &= 1 \times 1 + e^{-\frac{i\pi}{4}} \times e^{\frac{i\pi}{4}} + e^{-\frac{i\pi}{2}} \times e^{\frac{i\pi}{2}} \\ &\quad + e^{-\frac{i3\pi}{4}} \times e^{\frac{i3\pi}{4}} = 4 \\ \boldsymbol{q}(0.5) \cdot \boldsymbol{q}(1.5) &= 1 \times 1 + e^{\frac{i3\pi}{4}} \times e^{-\frac{i\pi}{4}} + e^{-\frac{i\pi}{2}} \times e^{-\frac{i\pi}{2}} \\ &\quad + e^{\frac{i\pi}{4}} \times e^{-\frac{i3\pi}{4}} = 0 \\ \boldsymbol{q}(0.5) \cdot \boldsymbol{q}(2.5) &= 1 \times 1 + e^{-\frac{i3\pi}{4}} \times e^{-\frac{i\pi}{4}} + e^{\frac{i\pi}{2}} \times e^{-\frac{i\pi}{2}} \\ &\quad + e^{-\frac{i\pi}{4}} \times e^{-\frac{i3\pi}{4}} = 0 \\ \boldsymbol{q}(0.5) \cdot \boldsymbol{q}(3.5) &= 1 \times 1 + e^{\frac{i\pi}{4}} \times e^{-\frac{i\pi}{4}} + e^{\frac{i\pi}{2}} \times e^{-\frac{i\pi}{2}} \\ &\quad + e^{-\frac{i3\pi}{4}} \times e^{-\frac{i3\pi}{4}} = 0 \\ \boldsymbol{q}(1.5) \cdot \boldsymbol{q}(2.5) &= 1 \times 1 + e^{-\frac{i3\pi}{4}} \times e^{-\frac{i3\pi}{4}} + e^{\frac{i\pi}{2}} \times e^{\frac{i\pi}{2}} \\ &\quad + e^{-\frac{i\pi}{4}} \times e^{-\frac{i\pi}{4}} = 0 \\ \boldsymbol{q}(1.5) \cdot \boldsymbol{q}(3.5) &= 1 \times 1 + e^{-\frac{i3\pi}{4}} \times e^{-\frac{i\pi}{4}} + e^{-\frac{i\pi}{2}} \times e^{\frac{i\pi}{2}} \\ &\quad + e^{-\frac{i3\pi}{4}} \times e^{-\frac{i\pi}{4}} = 0\end{aligned} \quad (29)$$

$$\begin{aligned}\boldsymbol{q}(2.5) \cdot \boldsymbol{q}(3.5) &= 1 \times 1 + e^{-\frac{i\pi}{4}} \times e^{\frac{i3\pi}{4}} + e^{-\frac{i\pi}{2}} \times e^{-\frac{i\pi}{2}} \\ &\quad + e^{-\frac{i3\pi}{4}} \times e^{\frac{i\pi}{4}} = 0\end{aligned}$$

When the selected frequencies are (-(*N*-1)/2: (*N*-1)/2), the four new perpendicular orthogonal vectors are listed in (30).

$$\begin{aligned}\boldsymbol{q}(0.5) &= \left[e^{-\frac{i\pi}{8}}, e^{-\frac{i3\pi}{8}}, e^{-\frac{i5\pi}{8}}, e^{-\frac{i7\pi}{8}}\right]^T \\ \boldsymbol{q}(1.5) &= \left[e^{-\frac{i3\pi}{8}}, e^{\frac{i7\pi}{8}}, e^{\frac{i\pi}{8}}, e^{-\frac{i5\pi}{8}}\right]^T \\ \boldsymbol{q}(2.5) &= \left[e^{-\frac{i5\pi}{8}}, e^{\frac{i\pi}{8}}, e^{\frac{i7\pi}{8}}, e^{-\frac{i3\pi}{8}}\right]^T \\ \boldsymbol{q}(3.5) &= \left[e^{-\frac{i7\pi}{8}}, e^{-\frac{i5\pi}{8}}, e^{-\frac{i3\pi}{8}}, e^{-\frac{i\pi}{8}}\right]^T\end{aligned} \quad (30)$$

The inner product of arbitrary two perpendicular vectors is zero, and the length of each perpendicular vector is 2, as shown in (31).

$$\begin{aligned}\boldsymbol{q}(0.5) \cdot \boldsymbol{q}(0.5) &= e^{\frac{i\pi}{8}} \times e^{-\frac{i\pi}{8}} + e^{\frac{i3\pi}{8}} \times e^{-\frac{i3\pi}{8}} + e^{\frac{i5\pi}{8}} \\ &\quad \times e^{-\frac{i5\pi}{8}} + e^{\frac{i7\pi}{8}} \times e^{-\frac{i7\pi}{8}} = 4 \\ \boldsymbol{q}(1.5) \cdot \boldsymbol{q}(1.5) &= e^{\frac{i3\pi}{8}} \times e^{-\frac{i3\pi}{8}} + e^{-\frac{i7\pi}{8}} \times e^{\frac{i7\pi}{8}} + e^{-\frac{i\pi}{8}} \\ &\quad \times e^{\frac{i\pi}{8}} + e^{\frac{i5\pi}{8}} \times e^{-\frac{i5\pi}{8}} = 4 \\ \boldsymbol{q}(2.5) \cdot \boldsymbol{q}(2.5) &= e^{\frac{i5\pi}{8}} \times e^{-\frac{i5\pi}{8}} + e^{-\frac{i\pi}{8}} \times e^{\frac{i\pi}{8}} + e^{-\frac{i7\pi}{8}} \\ &\quad \times e^{\frac{i7\pi}{8}} + e^{\frac{i3\pi}{8}} \times e^{-\frac{i3\pi}{8}} = 4 \\ \boldsymbol{q}(3.5) \cdot \boldsymbol{q}(3.5) &= e^{\frac{i7\pi}{8}} \times e^{-\frac{i7\pi}{8}} + e^{\frac{i5\pi}{8}} \times e^{-\frac{i5\pi}{8}} + e^{\frac{i3\pi}{8}} \\ &\quad \times e^{-\frac{i3\pi}{8}} + e^{\frac{i\pi}{8}} \times e^{-\frac{i\pi}{8}} = 4 \\ \boldsymbol{q}(0.5) \cdot \boldsymbol{q}(1.5) &= e^{\frac{i3\pi}{8}} \times e^{-\frac{i\pi}{8}} + e^{-\frac{i7\pi}{8}} \times e^{-\frac{i3\pi}{8}} + e^{-\frac{i\pi}{8}} \\ &\quad \times e^{-\frac{i5\pi}{8}} + e^{\frac{i5\pi}{8}} \times e^{-\frac{i7\pi}{8}} = 0 \\ \boldsymbol{q}(0.5) \cdot \boldsymbol{q}(2.5) &= e^{\frac{i5\pi}{8}} \times e^{-\frac{i\pi}{8}} + e^{-\frac{i\pi}{8}} \times e^{-\frac{i3\pi}{8}} + e^{-\frac{i7\pi}{8}} \\ &\quad \times e^{-\frac{i5\pi}{8}} + e^{\frac{i3\pi}{8}} \times e^{-\frac{i7\pi}{8}} = 0 \\ \boldsymbol{q}(0.5) \cdot \boldsymbol{q}(3.5) &= e^{\frac{i7\pi}{8}} \times e^{-\frac{i\pi}{8}} + e^{\frac{i5\pi}{8}} \times e^{-\frac{i3\pi}{8}} + e^{\frac{i3\pi}{8}} \\ &\quad \times e^{-\frac{i5\pi}{8}} + e^{\frac{i\pi}{8}} \times e^{-\frac{i7\pi}{8}} = 0 \\ \boldsymbol{q}(1.5) \cdot \boldsymbol{q}(2.5) &= e^{\frac{i5\pi}{8}} \times e^{-\frac{i3\pi}{8}} + e^{-\frac{i\pi}{8}} \times e^{\frac{i7\pi}{8}} + e^{-\frac{i7\pi}{8}} \\ &\quad \times e^{\frac{i\pi}{8}} + e^{\frac{i3\pi}{8}} \times e^{-\frac{i5\pi}{8}} = 0 \\ \boldsymbol{q}(1.5) \cdot \boldsymbol{q}(3.5) &= e^{\frac{i7\pi}{8}} \times e^{-\frac{i3\pi}{8}} + e^{\frac{i5\pi}{8}} \times e^{\frac{i7\pi}{8}} + e^{\frac{i3\pi}{8}} \\ &\quad \times e^{\frac{i\pi}{8}} + e^{\frac{i\pi}{8}} \times e^{-\frac{i5\pi}{8}} = 0 \\ \boldsymbol{q}(2.5) \cdot \boldsymbol{q}(3.5) &= e^{\frac{i7\pi}{8}} \times e^{-\frac{i5\pi}{8}} + e^{\frac{i5\pi}{8}} \times e^{\frac{i\pi}{8}} + e^{\frac{i3\pi}{8}} \\ &\quad \times e^{\frac{i7\pi}{8}} + e^{\frac{i\pi}{8}} \times e^{-\frac{i3\pi}{8}} = 0\end{aligned} \quad (31)$$

*E. The corrected SDFT*

According to the discussion above, the corrected even SDFT is defined as

$$X_s(m) = \sum_{n=-(N-1)/2}^{(N-1)/2} x(n + (N-1)/2) e^{-i2\pi mn/N}. \quad (32)$$

When the selected frequencies *m* are (-*N*/2: *N*/2-1), the transform matrix of the corrected SDFT is

$$\mathcal{F} = \begin{bmatrix} W^{(-\frac{N}{2})(-\frac{N-1}{2})} & W^{(-\frac{N}{2})(1-\frac{N-1}{2})} \\ W^{(1-\frac{N}{2})(-\frac{N-1}{2})} & W^{(1-\frac{N}{2})(1-\frac{N-1}{2})} \\ W^{(2-\frac{N}{2})(-\frac{N-1}{2})} & W^{(2-\frac{N}{2})(1-\frac{N-1}{2})} \\ \vdots & \vdots \\ W^{(\frac{N}{2}-1)(-\frac{N-1}{2})} & W^{(\frac{N}{2}-1)(1-\frac{N-1}{2})} \end{bmatrix}$$

$$\begin{bmatrix} W^{(-\frac{N}{2})(2-\frac{N-1}{2})} & \cdots & W^{(-\frac{N}{2})(\frac{N-1}{2})} \\ W^{(1-\frac{N}{2})(2-\frac{N-1}{2})} & \cdots & W^{(1-\frac{N}{2})(\frac{N-1}{2})} \\ W^{(2-\frac{N}{2})(2-\frac{N-1}{2})} & \cdots & W^{(2-\frac{N}{2})(\frac{N-1}{2})} \\ \vdots & \ddots & \vdots \\ W^{(\frac{N}{2}-1)(2-\frac{N-1}{2})} & \cdots & W^{(\frac{N}{2}-1)(\frac{N-1}{2})} \end{bmatrix} \quad (33)$$

In which the definition of $W^{mn}$ refers to (34).

$$W^{mn} = \exp(-i2\pi mn/N). \quad (34)$$

Where $m$ in (32) is not necessarily an integer, we choose integers to preserve frequency zero and take the FFT to realize fast SDFT. In this way, the corrected SDFT is inter-convertible with ODFT. According to (6) and (32), the interconverting formula of SDFT ($X_s$) and ODFT ($X_o$) is

$$X_s(m) = X_o(m)e^{-i\pi m(N-1)/N}. \quad (35)$$

The advisable frequencies of the corrected SDFT are (-(N-1)/2: (N-1)/2). The transform matrix of this form is (36). There is no fast algorithm for this form because the selected frequencies are different from the ODFT. As we have introduced the correction, SDFT below is the corrected even SDFT, which will not be stressed in the following paragraph.

$$\mathcal{F} = \begin{bmatrix} W^{(-\frac{N-1}{2})(-\frac{N-1}{2})} & W^{(-\frac{N-1}{2})(1-\frac{N-1}{2})} \\ W^{(1-\frac{N-1}{2})(-\frac{N-1}{2})} & W^{(1-\frac{N-1}{2})(1-\frac{N-1}{2})} \\ W^{(2-\frac{N-1}{2})(-\frac{N-1}{2})} & W^{(2-\frac{N-1}{2})(1-\frac{N-1}{2})} \\ \vdots & \vdots \\ W^{(\frac{N-1}{2})(-\frac{N-1}{2})} & W^{(\frac{N-1}{2})(1-\frac{N-1}{2})} \end{bmatrix}$$

$$\begin{bmatrix} W^{(-\frac{N-1}{2})(2-\frac{N-1}{2})} & \cdots & W^{(-\frac{N-1}{2})(\frac{N-1}{2})} \\ W^{(1-\frac{N-1}{2})(2-\frac{N-1}{2})} & \cdots & W^{(1-\frac{N-1}{2})(\frac{N-1}{2})} \\ W^{(2-\frac{N-1}{2})(2-\frac{N-1}{2})} & \cdots & W^{(2-\frac{N-1}{2})(\frac{N-1}{2})} \\ \vdots & \ddots & \vdots \\ W^{(\frac{N-1}{2})(2-\frac{N-1}{2})} & \cdots & W^{(\frac{N-1}{2})(\frac{N-1}{2})} \end{bmatrix} \quad (36)$$

*F. Inverse transform matrix*

Assuming the DFT matrix's orthogonal basis is standard, the inverse matrix is the forward matrix's Hermitian matrix, as shown in (37).

$$\mathcal{F}^{-1} = \mathcal{F}^{H} \quad (37)$$

If the forward transform matrix is (33) or (36), then the inverse transform matrix is

$$\mathcal{F}^{-1} = \frac{1}{N}\mathcal{F}^{H}. \quad (38)$$

*G. The spectrum of the three DFTs*

Assuming the FT of a continue signal $x(t)$ is $X(f)$, one obtains $N$ equidistant samples with sampling frequency $f_s$. According to the theory of convolution, the DFT of the $N$ samples is

$$X_D(f) = X(f) \otimes S(f) \otimes W(f). \quad (39)$$

Where "$\otimes$" represents convolution, $S(f)$ is the FT of the sampling function, and $W(f)$ is the FT of the default window. The sampling time interval is $t \in (-N/f_s/2, N/f_s/2)$. The default window of SDFT is

$$w(t) = \begin{cases} 1, & |t| < N/(2f_s) \\ 0, & |t| \geq N/(2f_s) \end{cases}. \quad (40)$$

The Fourier transform of the default window is

$$W(f) = \int_{-\infty}^{\infty} w(t)e^{-i2\pi ft}dt = \frac{N}{f_s}\text{sinc}(Nf/f_s). \quad (41)$$

From (41), we can see the FT of the default window is a real even function.

*1) ODFT*

For the sake of description, a single tone with a rectangular window is discussed. According to (39), the ODFT of the single tone is

$$X(f) = \sum_{j=-\infty}^{\infty} \frac{NA}{2} \begin{bmatrix} \text{sinc}\left(\frac{N}{f_s}(f - f_0 + jf_s)\right) e^{i(c_1(f - f_0 + jf_s) + \varphi)} \\ +\text{sinc}\left(\frac{N}{f_s}(f + f_0 + jf_s)\right) e^{-i(c_1(f + f_0 + jf_s) + \varphi)} \end{bmatrix} \quad (42)$$

Where $N$ is signal length, $f_s$ is the sampling frequency, $A$ is the amplitude of the single tone, $f_0$ is the single tone frequency, and $\varphi$ is the zero-point phase of the single tone. The value of constant $c_1$ is $-(N-1)\pi/N$. When $N$ is large enough, the ODFT of the single-tone is simplified as

$$X(f) \approx \frac{NA}{2}\text{sinc}\left(\frac{N}{f_s}(f - f_0)\right) e^{i(c_1(f - f_0) + \varphi)} \\ + \frac{NA}{2}\text{sinc}\left(\frac{N}{f_s}(f + f_0)\right) e^{-i(c_1(f + f_0) + \varphi)}. \quad (43)$$

*2) Uncorrected SDFT*

The uncorrected SDFT of the single tone is

$$X(f) = \frac{NA}{2}\sum_{j=-\infty}^{\infty}\text{sinc}\left(\frac{N}{f_s}(f - f_0 + jf_s)\right) e^{i(c_2(f - f_0 + jf_s) + \varphi)} + \\ \frac{NA}{2}\sum_{j=-\infty}^{\infty}\text{sinc}\left(\frac{N}{f_s}(f + f_0 + jf_s)\right) e^{-i(c_2(f + f_0 + jf_s) + \varphi)} \quad (44)$$

Where the value of constant $c_2$ is $\pi/N$. When $N$ is large enough, then the SDFT of a harmonic signal is simplified as

$$X(f) \approx \frac{NA}{2}\text{sinc}\left(\frac{N}{f_s}(f - f_0)\right) e^{i(c_2(f - f_0) + \varphi)} \\ + \frac{NA}{2}\text{sinc}\left(\frac{N}{f_s}(f + f_0)\right) e^{-i(c_2(f + f_0) + \varphi)}. \quad (45)$$

*3) Corrected SDFT*

The corrected SDFT of the single tone is

$$X(f) = \sum_{j=-\infty}^{\infty} \frac{NA}{2}(-1)^j \text{sinc}\left(\frac{N}{f_s}(f - f_0 + jf_s)\right) e^{i\varphi} \\ + \sum_{j=-\infty}^{\infty} \frac{NA}{2}(-1)^j \text{sinc}\left(\frac{N}{f_s}(f + f_0 + jf_s)\right) e^{-i\varphi}. \quad (46)$$

When $N$ is large enough, then the corrected SDFT of a harmonic signal is simplified as

$$X(f) \approx \frac{NA}{2}\text{sinc}\left(\frac{N}{f_s}(f - f_0)\right) e^{i\varphi} \quad (47)$$

$$+\frac{NA}{2}\text{sinc}\left(\frac{N}{f_s}(f+f_0)\right)e^{-i\varphi}.$$

The phase $\varphi$ in (42), (44), and (46) is the phase of zero-point. For ODFT, zero-point is at the head of the signal. For SDFT, zero-point is in the middle of the signal. Due to the different definitions of the zero-point, the zero-point phase is also different. Assuming the zero-point phase of the ODFT spectrum is $\varphi$, then the zero-point phase of the corrected SDFT spectrum is $\pi(N-1)f/f_s+\varphi$.

### H. Comparison of discrete Rectangular window

In order to show the superiority of the corrected SDFT, a discrete rectangular window is discussed. In which the length of the window is 20. According to (39), three discrete-time Fourier transforms (Ordinary DTFT, uncorrected symmetric DTFT, and corrected symmetric DTFT) of the rectangular window can be written as:

$$X_D(f)=\sum_{j=-\infty}^{\infty}N\text{sinc}\left(\frac{N}{f_s}(f+jf_s)\right)e^{ic_1(f+jf_s)} \quad (48)$$

$$X_D(f)=\sum_{j=-\infty}^{\infty}N\text{sinc}\left(\frac{N}{f_s}(f+jf_s)\right)e^{ic_2(f+jf_s)} \quad (49)$$

$$X_D(f)=\sum_{j=-\infty}^{\infty}(-1)^j N\text{sinc}\left(\frac{N}{f_s}(f+jf_s)\right) \quad (50)$$

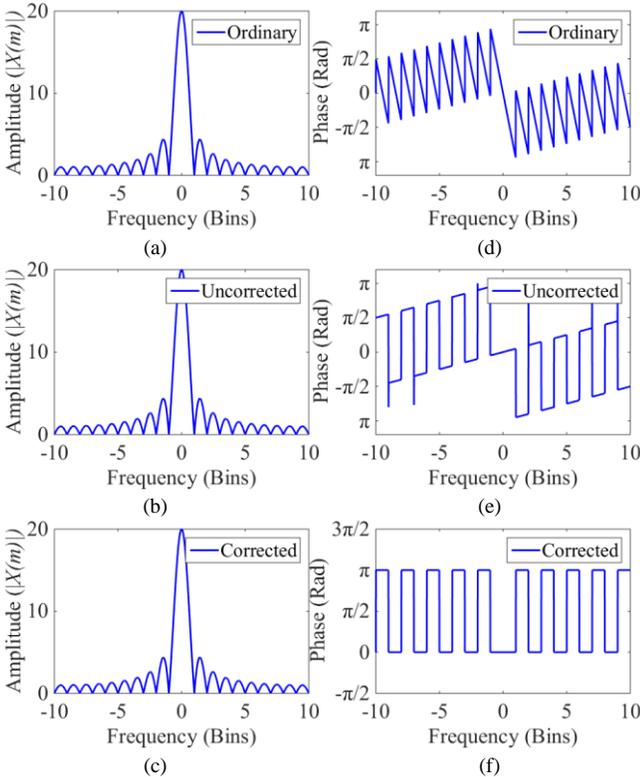

Fig. 4. Local DTFT spectra of rectangular window (window length $N$=20). (a), the amplitude spectra of ordinary DTFT; (b), the amplitude spectra of symmetric DTFT; (c), the amplitude spectra of corrected symmetric DTFT; (d), the phase spectra of ordinary DTFT; (e), the phase spectra of symmetric DTFT; (f), the phase spectra of corrected symmetric DTFT.

Three DTFT spectrums of the rectangular window are plotted in Fig. 4, in which signal length $N$ is 20 [26]. The amplitude spectrum is the same, whereas the phase spectrum is different. The phase spectrum of ordinary DTFT is a linear function of frequency, and the phase spectrum of corrected symmetric DTFT is only 0 or $\pi$. The phase spectrum of corrected symmetric DTFT can be converted to 0 by Euler's formula.

## IV. THE UNIQUE PROPERTY OF SDFT

DFT has many properties, and this study will not introduce them one by one. Property that both ODFT and SDFT have is not in the scope of this study. This study only introduces the properties that SDFT has, but ODFT does not have. For example, the symmetry properties and the integral properties. These unique properties of SDFT are the important reasons why the author recommends SDFT.

### A. Symmetry properties of SDFT

Assuming the FT of function $x(t)$ is $X(f)$. If $x(t)$ is real and even symmetry, then $X(f)$ is real and even symmetry. If $x(t)$ is real and odd symmetry, then $X(f)$ is imaginary and odd symmetry. However, ODFT does not have these symmetry properties.

Assuming the FT of a continue signal $x(t)$ is $X(f)$, one obtains $N$ equidistant samples with sampling frequency $f_s$. According to the theory of convolution, the SDFT of the $N$ samples is

$$X_D(f)=X(f)\otimes S(f)\otimes W(f). \quad (51)$$

Where "$\otimes$" represents convolution, $S(f)$ is the FT of the sampling function, and $W(f)$ is the FT of the default window. According to subsection A of section III, $S(f)$ is real and even for odd sampling and even sampling function. According to (41), $W(f)$ is real and even. If $X(f)$ is an odd and function, we have

$$X_D(-f)=X(-f)\otimes S(-f)\otimes W(-f)=-X_D(f). \quad (52)$$

If $X(f)$ is an even function, then we have

$$X_D(-f)=X(-f)\otimes S(-f)\otimes W(-f)=X_D(f). \quad (53)$$

Then we can conclude that SDFT has the same symmetry properties as that of FT.

### B. The time-domain integral property

The integral properties of FT can be divided into the time-domain integral property and frequency-domain integral property[27]. The time-domain integral property of FT can be summarized as follows: assuming $x(t)$ is an integrable function, and the FT of $x(t)$ is $X(f)$. The frequency-domain origin equals $x(t)$'s integral over all its time-domain, as shown in (54).

$$X(0)=\int_{-\infty}^{\infty}x(t)\,\mathrm{d}t \quad (54)$$

The integral property of FT is also applicable to both ODFT and SDFT, as shown in (55) and (56).

$$X(0)=\sum_{n=0}^{N-1}x(n) \quad (55)$$

$$X(0)=\sum_{n=-(N-1)/2}^{(N-1)/2}x(n) \quad (56)$$

The theoretical derivation of the above three formulas can be obtained by substituting frequency zero into its transform definition formula.

### C. The frequency-domain integral property of FT

The frequency-domain integration of FT is divided into real-part integration and imaginary-part integration. The real-part integral property can be summarized as follows: assuming $x(t)$ is an integrable function, and the FT of $x(t)$ is $X(f)$, the time-domain origin $x(0)$ equals the real-part integration over its frequency-domain, as shown in (57).

$$x(0) = \int_{-\infty}^{\infty} X(f) \, df \tag{57}$$

There are two ways to prove this formula. The simplest way is substituting time zero into the definition formula of inverse FT (IFT). The other way is introduced in the following paragraphs.

The real part of an FT spectrum is even symmetry, and the imaginary part is odd symmetry. What's more, according to (54), the frequency-domain origin is real-valued. Hence, the imaginary-part integral can be written as

$$J_{\text{im}} = \int_0^{\infty} X(f) \, df - \int_{-\infty}^0 X(f) \, df. \tag{58}$$

The imaginary-part integral property can be summarized as follows: assuming $x(t)$ is an integrable function, $h(t)$ is the Hilbert transform of $x(t)$, and the FT of $x(t)$ is $X(f)$; the imaginary-part integration over its frequency-domain equals $h(0)$, as shown in (59).

$$h(0) = \int_0^{\infty} X(f) \, df - \int_{-\infty}^0 X(f) \, df \tag{59}$$

*1) The real prat integral property*

Another way to prove the property of the real-part integral of FT is shown below. Assuming a continuous signal $x(t)$ has $k$ constituent frequencies, then $x(t)$ can be written as

$$x(t) = \sum_{i=0}^{k-1} x_i(t) = \sum_{i=0}^{k-1} A_i \cos(2\pi f_i t + \varphi_i). \tag{60}$$

Where $A_i$, $f_i$, and $\varphi_i$ are the amplitude, frequency, and phase of $x_i(t)$. The FT of the continuous signal is

$$X(f) = \sum_{i=0}^{k-1} X_i(f) = \sum_{i=0}^{k-1} \left( \frac{A_i}{2} \delta(f - f_i) e^{i\varphi_i} + \frac{A_i}{2} \delta(f + f_i) e^{-i\varphi_i} \right). \tag{61}$$

The integration of the real-part can be written as

$$J_{\text{real}} = \int_{-\infty}^{\infty} X(f) \, df. \tag{62}$$

Substituting (61) into (62), then we have:

$$J_{\text{real}} = \sum_{i=0}^{k-1} A_i \cos(\varphi_i) = x(0). \tag{63}$$

From (63), we can see the real-part integration over its frequency-domain equals $x(0)$.

*2) The imaginary prat integral property*

The proof of the integration of the imaginary part is as follows. Substituting Eq. (61) into Eq. (58), then we have

$$J_{\text{im}} = i \sum_{i=0}^{k-1} A_i \sin(\varphi_i) = i \sum_{i=0}^{k-1} A_i \cos(\varphi_i - \pi/2). \tag{64}$$

We can see that the integration of the imaginary part is equal to the origin of its Hilbert transform.

*D. Summation of the real part of odd SDFT*

In this section, the author discusses the sum of the real parts of odd SDFT. Assuming a discrete signal $x(n)$ has $k$ constituent frequency components, then $x(n)$ can be written as

$$x(n) = \sum_{i=0}^{k-1} x_i(n) = \sum_{i=0}^{k-1} A_i \cos(2\pi f_i n/f_s + \varphi_i). \tag{65}$$

Where $A_i$, $f_i$, and $\varphi_i$ are the amplitude, frequency, and phase, respectively. Assuming the SDFT of arbitrary constituent component $x_i(n)$ is $X_i(m)$, and the real-part summation over the frequency-domain is given by (66). In which $N$ is the length of the discrete signal.

$$S_{\text{re}}(i) = \sum_{m=-(N-1)/2}^{(N-1)/2} X_i(m) \tag{66}$$

According to the theory of convolution, the odd SDFT of the Fourier component $x_i(n)$ is

$$X_i(m) = \sum_{j=-\infty}^{\infty} \frac{NA_i}{2} \text{sinc}\left(m - \frac{f_i N}{f_s} + jN\right) e^{i\varphi_i} \\ + \sum_{j=-\infty}^{\infty} \frac{NA_i}{2} \text{sinc}\left(m + \frac{f_i N}{f_s} + jN\right) e^{-i\varphi_i}. \tag{67}$$

After simplification, $X_i(m)$ can be written as:

$$X_i(m) = \frac{NA_i}{2} \sum_{j=-\infty}^{\infty} X_i(m,j). \tag{68}$$

In which $X_i(m,j)$ can be written as:

$$X_i(m,j) = \text{sinc}\left(m - \frac{f_i N}{f_s} + jN\right) e^{i\varphi_i} \\ + \text{sinc}\left(m + \frac{f_i N}{f_s} + jN\right) e^{-i\varphi_i}. \tag{69}$$

If we take variable $p$ to replace $m+jN$, then we have:

$$X_i(m,j) = \text{sinc}\left(p - \frac{f_i N}{f_s}\right) e^{i\varphi_i} \\ + \text{sinc}\left(p + \frac{f_i N}{f_s}\right) e^{-i\varphi_i}. \tag{70}$$

Substituting (68) and (70) into (66), then we have:

$$S_{\text{re}}(i) = \frac{NA_i}{2} \sum_{h=-\infty}^{\infty} \text{sinc}\left(p - \frac{f_i N}{f_s}\right) e^{i\varphi_i} \\ + \frac{NA_i}{2} \sum_{h=-\infty}^{\infty} \text{sinc}\left(p + \frac{f_i N}{f_s}\right) e^{-i\varphi_i}. \tag{71}$$

Because $m$, $j$, and $N$ are integers, variable $p$ is an integer. For arbitrary real number $q$, we have

$$\sum_{h=-\infty}^{\infty} \text{sinc}(p - q) = 1. \tag{72}$$

Substituting (72) into (71), then we have:

$$S_{\text{re}}(i) = NA_i \cos(\varphi_i) = Nx_i(0). \tag{73}$$

According to the linearity of FT, the real part summation over the frequency-domain can be written as

$$S_{\text{re}} = \sum_{i=0}^{k-1} Nx_i(0) = Nx(0). \tag{74}$$

From (74), we can conclude that the real-parts summation of the odd SDFT spectrum is equal to $N$ times the time-domain origin.

*E. Sum of the imaginary-part of odd SDFT*

The real part of the SDFT spectrum is even symmetric, and the imaginary part of the SDFT spectrum is odd symmetric. In this section, the author discussed the sum of the imaginary part. According to (56), the frequency-domain origin is real-valued. That is to say, $X(0)$ is negligible when discussing the imaginary spectrum. Assuming signal length $N$ is an odd number, then the sum of the imaginary part is

$$S_{\text{im}} = \sum_{m=1}^{(N-1)/2} X(m) - \sum_{m=-(N-1)/2}^{-1} X(m). \quad (75)$$

To simply the derivation, only one constituent frequency $x_i(n)$ is discussed. Once the imaginary-part summation of $X_i(m)$ is obtained, and according to the linearity of SDFT, the imaginary-part summation of $X(m)$ can be deduced easily. Substituting (67) into (75), then we have

$$S_{\text{im}}(i) = \frac{NA_i}{2}e^{i\varphi_i}\gamma_i + \frac{NA_i}{2}e^{-i\varphi_i}\delta_i. \quad (76)$$

In which $\gamma_i$ and $\delta_i$ are only related to $N$, $f_s$, and $f_i$, as shown in (77) and (78).

$$\gamma_i = \sum_{m=1}^{\frac{(N-1)}{2}} \sum_{j=-\infty}^{\infty} \text{sinc}\left(m - \frac{f_i N}{f_s} + jN\right)$$
$$- \sum_{m=-(N-1)/2}^{-1} \sum_{j=-\infty}^{\infty} \text{sinc}\left(m - \frac{f_i N}{f_s} + jN\right), \quad (77)$$

$$\delta_i = \sum_{m=1}^{\frac{(N-1)}{2}} \sum_{j=-\infty}^{\infty} \text{sinc}\left(m + \frac{f_i N}{f_s} + jN\right)$$
$$- \sum_{m=-(N-1)/2}^{-1} \sum_{j=-\infty}^{\infty} \text{sinc}\left(m + \frac{f_i N}{f_s} + jN\right). \quad (78)$$

According to the symmetry property of the SDFT spectrum, we have $\gamma_i = -\delta_i$, (76) is simplified as

$$S_{\text{im}}(i) = \gamma_i \left(\frac{NA_i}{2}e^{i\varphi_i} - \frac{NA_i}{2}e^{-i\varphi_i}\right) = i\gamma_i NA_i \sin(\varphi_i). \quad (79)$$

According to the linearity of SDFT, the sum of the imaginary parts becomes

$$S_{\text{im}} = i \sum_{i=0}^{k-1} \gamma_i NA_i \cos(\varphi_i - \pi/2). \quad (80)$$

Although each component's amplitude is scaled due to the spectrum leakage effect, the phase is correct. The scale factor ($\gamma_i$) changes with frequency. Even SDFT has a similar integral property. The derivation is similar and neglected. A summary of the integral property is listed in Table II. In which parameters $\alpha$, $\beta$, and $\gamma$ have the translation invariance and rotational invariance property. They only relate to the sampling frequency, signal length ($N$), and signal frequency.

TABLE II
THE INTEGRAL PROPERTY OF DIFFERENT TRANSFROMS

| Transform | Real part integration | Imaginary part integration |
|---|---|---|
| FT | $x(0)$ | $ih(0)$ |
| ODFT | - | - |
| Odd SDFT | $Nx(0)$ | $i\sum_{i=0}^{k-1}\gamma_i NA_i \sin(\varphi_i)$ |
| Even SDFT | $\sum_{i=0}^{k-1}\alpha_i NA_i \cos(\varphi_i)$ | $i\sum_{i=0}^{k-1}\beta_i NA_i \sin(\varphi_i)$ |

Where $N$ is signal length. Parameters $\alpha$, $\beta$, and $\gamma$ have the translation invariance and rotational invariance property.

## V. ZERO PADDING TECHNIQUE

Zero-padding is a technique defined as appending zero values to the weighted samples prior to the DFT calculation. The appended zero values are treated as additional samples collected at the same rate and therefore extend the measurement time [28].

Note that extending the data with zeros and computing a longer DFT increases the number of points in the frequency-domain but does not break the basic restriction, nor does it alter aliasing effects [29]. Resolution limits are determined by the observation interval and the sampling frequency [29]. No amount of zero paddings can overcome these fundamental limits, and the spectrum parameters, such as signal-to-noise ratio level and the spectral leakage level, remain unchanged [28].

### A. Time-domain zero padding

Time-domain zero-padding for ODFT is shown in (81), in which it pads a large number of zeros at the end of the weighted samples.

$$x_{wp}(n) = \begin{cases} x_w(n) & 0 \leq n < N \\ 0 & N \leq n < MN \end{cases} \quad (81)$$

Where $x_w$ is the weighted samples, and $(M-1)N$ zeros are padded. Accordingly, the discrete spectrum extends as well. Instead of $N$ spectrum samples, $M*N$ spectrum samples of the same spectrum are made available. The distance between arbitrary two spectrum samples is $1/M$ bins, which is adjustable.

Zero-padding for SDFT spectrum calculation is shown in (82) [26]. In which it pads the same amount of zeros at the two ends.

$$x_{wp}(n) = \begin{cases} 0 & -M(N-1)/2 \leq n < -(N-1)/2 \\ x_w(n) & -(N-1)/2 \leq n \leq (N-1)/2 \\ 0 & (N-1)/2 < n \leq M(N-1)/2 \end{cases} \quad (82)$$

### B. Frequency domain zero padding

In the complex number field $\mathbb{C}$, the time-domain and frequency-domain are relative to each other. Both ODFT and SDFT are orthogonal transforms. Hence, time-domain samples and frequency-domain samples are equivalent to each other. Theoretically, if there is a time-domain zero-padding technique, there is a frequency-domain zero-padding technique.

By applying the inverse discrete Fourier transform (IDFT), the time-domain samples can be reconstructed. Theoretically, the time-domain samples would be significantly increased if one pads a large number of zeros in the frequency-domain. One application of frequency-domain zero-padding is the interpolation technique. The interpolation process can be divided into three steps. First, perform DFT on the input signal. Second, pad zeros for the spectrum samples. Last, perform IDFT. The real part of the inverse transformation is the output signal.

The author has tried frequency-domain zero-padding for both ODFT and SDFT according to (81) and (82), respectively, and found frequency-domain zero-padding only suitable for SDFT. That is to say, SDFT can be used for interpolation, but ODFT cannot. More information can be found in subsection *D* of section V.

### C. Discrete frequency Fourier transform

The definition of discrete frequency Fourier transform (DFFT) is

$$x(n) = \frac{1}{N} \sum_{m=-(N-1)/2}^{(N-1)/2} X(m)\, e^{i2\pi mn/N} \tag{83}$$

Where $X(m)$ is the discrete spectrum samples obtained by SDFT, and $N$ is the signal length. One difference between DFFT and inverse SDFT is the value range of $n$, in which the value range of inverse SDFT is $(-(N-1)/2:1:(N-1)/2)$, whereas the value range of DFFT is $\{n \in \mathbf{R}| -\infty < n < \infty\}$.

When the time-domain zero-padding parameter ($M$) tends to infinity, we obtain a period spectrum of the discrete-time Fourier transform (DTFT) by performing DFT. When it comes to the frequency-domain, assuming the zero-padding parameter ($M$) tends to infinity, a period of the reconstructed DFFT samples can be obtained by performing inverse SDFT.

TABLE III
THE OVERSHOOT AND UNDERSHOOT OF SQUARE WAVE

| $k$ | 11 | 41 | 501 | 100001 |
|---|---|---|---|---|
| Overshoot | 0.122486 | 0.135513 | 0.139830 | 0.140208 |
| Undershoot | -0.156931 | -0.144809 | -0.140590 | -0.140212 |

### D. The Gibbs phenomenon of SDFT

The Gibbs phenomenon describes a peculiar phenomenon of the Fourier series. The Fourier series of a piecewise continuously differentiable periodic function behaves at a jump discontinuity [30]. The $n^{\text{th}}$ partial sum of the Fourier series has large oscillations near the jump, which might increase the maximum of the partial sum above that of the function itself. The overshoot does not die out as $n$ increases but approaches a finite limit [30].

The author finds a similar phenomenon when reconstructing the time-domain with the $N$ spectrum samples. Without loss of generality, we may take the square wave as an example. The discrete square wave is

$$x(n) = \begin{cases} 0 & 1 \le n \le k \\ 1 & k < n \le 2k \\ 0 & 2k < n \le 3k \end{cases} \tag{84}$$

Both ODFT and SDFT spectrums are selected to reconstructing the time-domain signal. The frequency-domain zero-padding for the ODFT spectrum and SDFT spectrum is (81) and (82).

In this experiment, a series of square waves are simulated, in which the value of $k$ ranges from 11 to 100001. The output results are plot in Fig. 5, where the left side is the output results of ODFT based interpolation, and the right side is the output results of SDFT based interpolation. The interpolation error based on inverse ODFT is large, whereas the interpolation error based on inverse SDFT is small. Hence, ODFT is not suitable for DFFT.

The DFFT overshoots and undershoots at a jump discontinuity, and it does not die out. As signal length increases, the overshoot and undershoot are converge to a constant. Generally, at any jump point with a jump of $a$, the DFFT will overshoot this jump by approximately $0.140210a$ at one end and undershoot it by the same amount at the other end. The maximum overshoots and undershoots of the above experiment are listed in Table III. We can see that overshoot and undershoot both approximate 0.140210.

The above constant is different from the *Wilbraham-Gibbs Constant* [31] and needs further study. Due to the length limitation of this manuscript, the constant will not be explained further.

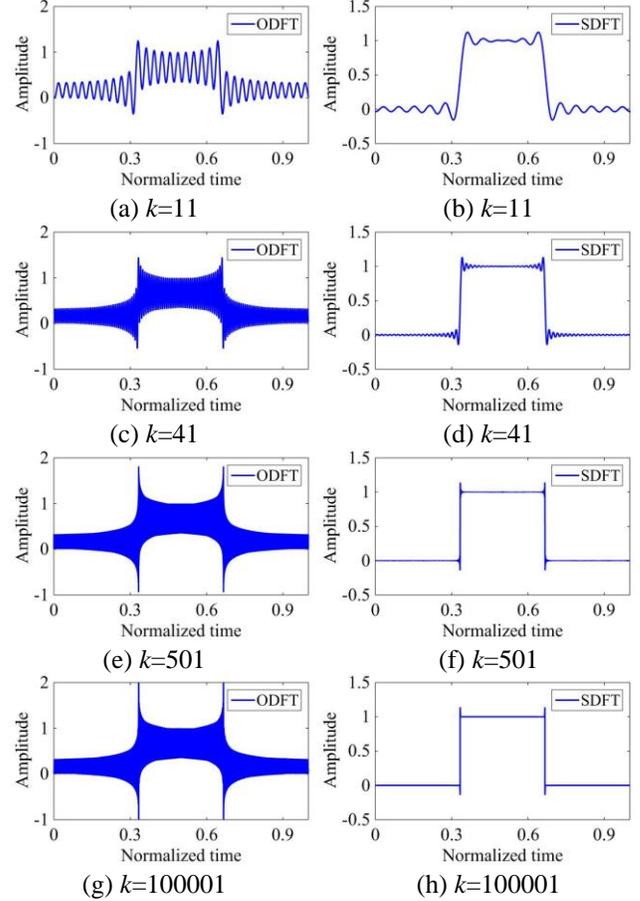

(a) $k=11$  (b) $k=11$
(c) $k=41$  (d) $k=41$
(e) $k=501$  (f) $k=501$
(g) $k=100001$  (h) $k=100001$

Fig. 5. The interpolated signal that based on time-domain zero-padding and inverse DFT. The zero-padding parameter of $M$ is 11. The left side is the output results of ODFT based interpolation. The right side is the output results of SDFT based interpolation.

## VI. DISCUSSION AND CONCLUSIONS

SDFT and ODFT are orthogonal transforms. The two DFTs' amplitude spectrum is the same in one-dimensional transform, whereas the two DFTs' phase spectrum is different. If the SDFT spectrum is chosen to replace the ODFT spectrum, it will not affect the application based on the amplitude spectrum. However, it may have a huge impact on applications based on phase spectrum. If one wants to get better results in applications based on phase spectrum, it would be a good attempt to replace ODFT with SDFT. It is hard to predict the output results, and one has to look for his fortune.

### A. Reasons for choosing SDFT

The author recommends SDFT based on the following six reasons. Given that SDFT has more FT properties than ODFT, the author believes SDFT will be widely applied in the future.

*1) Symmetry in the time-domain*

The time-domain of FT and SDFT are both symmetric to zero. However, the time-domain of ODFT is asymmetric to zero.

*2) Conjugate property*

Assuming the FT of a continue signal $x(t)$ is $X(f)$, if one turns the signal head around, one gets

$$x'(t) = x(-t). \tag{85}$$

Assuming the FT of the continue signal *x*'(*t*) is *X*'(*f*), then he may find that *X*(*f*) and *X*'(*f*) compose a conjugate pair. If one turns a discrete sequence head around, SDFT gets a pair of the conjugate spectrum, whereas ODFT gets two completely different spectra.

*3) Symmetry properties*

The symmetry properties of FT can be simplified as: assuming the FT of function *x*(*t*) is *X*(*f*); if *x*(*t*) is a purely real and even function, then *X*(*f*) is a purely real and even function; If *x*(*t*) is a purely real and odd function, then *X*(*f*) is a purely imaginary and odd function. SDFT has the same symmetry properties, whereas ODFT does not have these symmetry properties.

*4) Frequency domain integral properties*

The frequency-domain integral properties of FT can be summarized as follows: assuming *x*(*t*) is an integrable function, the FT of *x*(*t*) is *X*(*f*), and the Hilbert transform (HT) of *x*(*t*) is *h*(*t*); the time-domain origin *x*(0) equals the real-part integration over its frequency-domain; The origin of HT *h*(0) equals the imaginary-part integration over its frequency-domain. The integral property of SDFT is slightly different from FT due to the spectrum leakage effects, whereas ODFT does not have the frequency-domain integral properties.

*5) Interpolation property*

According to FT's definition, in the complex number field $\mathbb{C}$, the time-domain and frequency-domain are relative. Hence, the frequency-domain spectrum can describe time-domain signals and vice versa. DFT is the numerical implementation of FT. Theoretically, it has the same characteristics. The zero-padding technique in the time-domain can be used to obtaining more spectrum samples. That is to say, zero-padding can be used for frequency-domain interpolation. Hence, the zero-padding technique in the frequency-domain can be used to obtaining more time-domain samples. That is to say, zero-padding can be used for time-domain interpolation.

It has been verified that the time-domain zero-padding is suitable for both ODFT and SDFT. However, the frequency-domain zero-padding is only suitable for SDFT. Though the output results show a similar phenomenon to the Gibbs phenomenon, the output result is acceptable.

*6) Noether's theorem*

According to Noether's theorem [20], symmetry seems to be the prerequisite of a differentiable physical system with the conservation law.

*B. Reasons for choosing odd SDFT*

In this paragraph, the author raises a new issue: should the signal length be odd or even when performing DFT. The author's answer is "odd." However, in reality, scientists and engineers are accustomed to using even-numbered signals. That is the reason why this issue is important. Based on the following four reasons, the author recommends using an odd number of samples when performing SDFT.

*1) The centered spectrum*

The fftshift function shifts the zero-frequency component to the center of the spectrum. When the signal length is odd, the zero-frequency is in the middle of the spectrum. However, the zero-frequency is approximately in the middle of the spectrum when the signal length is even. Strictly speaking, the shifted spectrum is not symmetrical to zero. Though the approximation error is small for long signals, the error is large for short signals. If the signal's default length is an odd number, the asymmetry mentioned above will disappear.

*2) Zero-point phase*

The frequency spectrum is only related to signal amplitude, frequency, and zero-point phase. As one of the most important parameters, the zero-point phase is widely used in many applications. In SDFT, the zero-point locates in the middle of the discrete signal. When the signal length is odd, the zero-point is collected. When the signal length is even, the zero-point is missed. For more information, please refers to subsection *A* of Section II. The missing zero-point may cause a lot of trouble.

*3) Frequency domain period*

Assuming the sampling frequency of an odd sampling function and an even sampling function are $f_s$. In the time-domain, the period of both the odd sampling function and even sampling function is $1/f_s$. In the frequency-domain, the period of the two sampling functions is different. More specifically, the odd sampling function period is $f_s$, whereas the even sampling function is $2f_s$. It is abnormal that the periodic relationship between the time-domain and frequency-domain of even sampling function changes.

*4) The integral property of SDFT*

FT's frequency-domain integral property can be simplified as: the time-domain origin *x*(0) equals the real-part integration over its frequency-domain. Though both odd SDFT and even SDFT have the real-part integral property, the integration is different. The real-part integration of even SDFT has an unpredictable scaling factor, whereas odd SDFT does not have.